\documentclass[11pt,twoside]{article}

\usepackage{asp2010}

\resetcounters

\begin{document}

\title{A view of the solar neighbourhood: the local population of planetary nebulae and their mimics}

\author{David J. Frew$^1$ and Quentin A. Parker$^{1,2}$
\affil{$^1$Department of Physics and Astronomy, Macquarie University, NSW 2109, Australia}
\affil{$^2$Australian Astronomical Observatory, PO Box 296, Epping, NSW 1710, Australia}
}

\begin{abstract}
We have, for the first time, compiled a nearly complete census of planetary nebulae (PNe) centred on the Sun. Our goal is the generation of an unbiased volume-limited sample, in order to answer some long-standing statistical questions regarding the overall population of Galactic disk PNe and their central stars.  Much improved discrimination of classical PNe from their mimics is now possible based on the wide variety of high-quality multiwavelength data sets that are now available.  However, we note that evidence is increasing that PNe are heterogeneous, and probably derived from multiple evolutionary scenarios.  We give some preliminary data on the relative frequencies of different types of PNe in the local Galactic disk.\\
\noindent{\bf Keywords.}\hspace{12pt} Stars: AGB and post-AGB -- planetary nebulae: general
\end{abstract}

\section{Introduction}
Planetary nebulae (PNe) are an important, albeit brief, evolutionary phase in the lifetimes of a significant fraction of Milky Way stars.  While their formation is broadly understood (e.g. Balick \& Frank 2002), the exact mechanism(s) required to manufacture the multitude of PN morphologies remains unclear, as does the ultimate fate of our own Sun.  To help answer these questions, we have compiled the most complete volume-limited census of PNe out to 2.0~kpc from the Sun ever compiled (Frew \& Parker 2006, Frew 2008), containing 210 PNe.  We have recently extended the census out to 3.0~kpc (Frew \& Parker 2010b) which doubles the number to over 420 PNe, though this enlarged sample is less complete at the faint end of the PN luminosity function (Ciardullo 2010). This number can be compared to the known Galactic population which currently totals nearly 3000 PNe (Frew \& Parker 2010; Jacoby et al. 2010). 

Volume-limited samples are fundamental in astronomy, but not easy to produce. The ability to generate this PN census rests, in large part, on the application of our new, empirical H$\alpha$ surface brightness -- radius (SB-$r$) relation which we have shown can provide distances accurate to 20--30$\%$ (Frew \& Parker 2006; Frew 2008). Our technique is currently the only statistical method that is applicable to the very faintest, senile PNe, which are selected against in extant radio surveys.  Such faint PN numerically dominate any volume-limited sample, so it is crucial to include them in order to generate an unbiased census which can then be  used to answer some long-standing statistical questions regarding the overall population of Galactic disk PNe and their central stars (CSPN).  Estimates of the local volume density, scale height, and total number of PN in the Galaxy also rest on having an accurate census of nearby nebulae (see Frew \& Parker 2007, for a discussion).

Surprisingly, and somewhat distressingly, there remains a lack of consensus over the precise definition of a PN, a situation that lingers even after several decades of intensive effort\footnote{During the 1967 IAU PN symposium, R. Minkowski responded to a query from D.S. Evans, stating: \emph{As to the question of how to define a planetary nebula, there is no better way than to accept any object in a catalogue of planetary nebulae if nobody has serious objections} (Osterbrock \& O'Dell 1968: 290).} (for a fuller discussion of this problem, see Frew \& Parker 2010a).  It is commonly argued that the `planetary nebula' moniker has a distinct physical meaning, and should be restricted to the ionized shell ejected at the end of the AGB phase, or by a common-envelope ejection (De Marco 2009, and references therein).  This is an important point, as in many symbiotic systems, which are often confused with PNe, the gas is thought to be donated by a companion giant, and not derive from the precursor of the white dwarf (WD) which is usually present in these systems (Corradi 2003).

Besides our development of the H$\alpha$ SB-$r$ relation, we have also refined a range of multiwavelength classification tools to weed out the many mimics that have contaminated both Galactic and extragalactic PN catalogues in the past.  We have recently used mid-IR and radio diagnostics to identify contaminants (Cohen et al. 2007, 2010).  Based on an overview of the literature, extensive experience in the compilation of the MASH catalogues (Parker et al. 2006; Miszalski et al. 2008), as well as insights from the solar neighbourhood PN census,  Frew \& Parker (2010a) provided a phenomenological definition, listing a summary of the various observable properties manifested by PNe.

\begin{figure}[h]
\begin{center}
\includegraphics[scale=0.475]{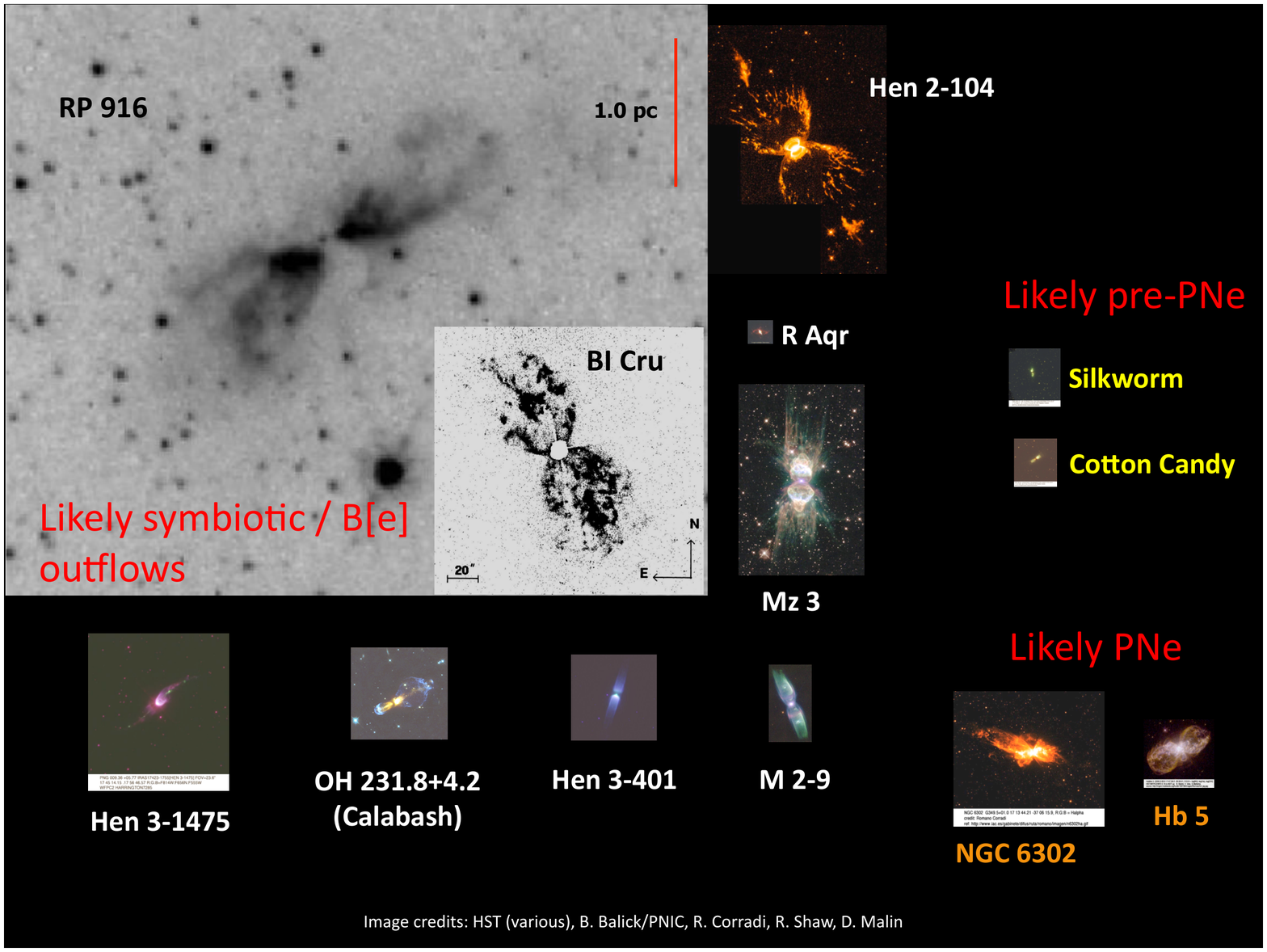}
\caption{A montage of selected ``PNe'' from the literature, all presented at the same physical size; the scale bar represents 1~pc.  We find that size appears to be a simple, but powerful discriminant between pre-PNe and most bona fide symbiotic outflows like BI~Cru and He~2-104.}
\label{fig:neb_size}
\end{center}
\end{figure}

\section{Identifying PNe}
A review of the history of PN surveys is presented by Parker \& Frew (2010, these proceedings).  PNe have been confused with many different types of objects, as diverse as HII regions, massive star ring nebulae, Herbig-Haro objects, B[e] stars, reflection nebulae, supernova remnants, symbiotic outflows, and galaxies.  We recommend that the following diagnostic criteria be examined, the exact combination depending on whether the candidate nebula is compact or extended.  These criteria have been rigorously applied to our local volume sample (as discussed by Frew \& Parker 2010a).  They include:\\

\noindent $\bullet$ Presence of a hot (bluish) star that is relatively faint compared to the nebular flux. Non-blue stars may be either reddened single stars or binary stars with a cooler companion; in the latter case, a UV excess can provide evidence for the true ionizing star (e.g. Frew et al. 2010b).  The properties of the ionizing star, if known, should be examined, including its evolutionary age and position in the HR diagram.\\
\noindent $\bullet$ Nebular morphology and the ionization structure, including the consistency of any ISM interaction with the proper-motion vector of the ionizing star.\\
\noindent $\bullet$ Nebular emission-line ratios, using diagnostic plots where applicable (see Frew \& Parker 2010a).\\
\noindent $\bullet$ Near-IR and mid-IR colours; use a range of diagnostic plots (Schmeja \& Kimeswenger 2001; Corradi et al. 2008; Cohen et al. 2007, 2010). \\
\noindent $\bullet$ Available time-domain photometric data to search for variability of the ionizing star, exhibited by close-binary CSPN or symbiotic stars.\\ 
\noindent $\bullet$ Strength of the radio and mid-IR flux densities (especially useful if a constraint on the distance is available).\\
\noindent $\bullet$ Systemic radial velocity of the nebula (does it differ from the RV of the ionizing star?) and the line width --- is the nebula expanding or is the line width consistent with an HII region or static ISM?  Alternatively, are  broad H$\alpha$ wings seen in compact objects, as this may be indicative of a symbiotic or B[e] star. \\
\noindent $\bullet$ Physical nebular diameter, which should be $\leq$5~pc, but a distance estimate is required.  While there is overlap in the sizes of PNe, Population~I ring nebulae, and resolved symbiotic outflows, we find bona-fide pre-PNe (Sahai et al. 2007; Szczerba et al. 2007) to be generally smaller than symbiotic outflows (Figure~\ref{fig:neb_size}).\\
\noindent $\bullet$ Ionized mass, which should be between 0.005 and 3$M_{\odot}$, but the distance needs to be known.\\
\noindent $\bullet$ Local environment. For example, YSOs, and T Tauri stars are usually associated with HII regions, molecular clouds, and areas of heavy obscuration or dark lanes.\\
\noindent $\bullet$ Galactic latitude. PNe have a larger scale height than HII regions, SNRs, and massive stars, and are more likely to be found away from the Galactic plane.\\
\noindent $\bullet$ Abundances of the nebular gas -- check for N enrichment, which would indicate either a PN, a nova shell, or massive star ejecta.

No one criterion is generally enough, so we use the overall body of evidence to define the status of a candidate nebula (see Frew \& Parker 2010a, for further details).  As an illustrative example of our multiwavelength approach, the emission nebula around the sdOB star PHL 932 is detailed by Frew et al. (2010a), and convincingly shown to be a HII region; it should be expunged from current catalogues of PNe. We have discarded another dozen objects from the 1.0 kpc census (e.g. Sh~2-174 and Abell~35, amongst others; Frew 2008).  The reader is also referred to Frew \& Parker (2006), Parker et al. (2006) and  Madsen et al. (2006) for other examples of nearby PN impostors.

\section{Do PNe Form a Heterogeneous Class?}\label{sec:hetero}
Even after removal of the obvious contaminants from PN catalogues, there is increasing evidence that the class itself is a `mixed bag'  (Frew \& Parker 2010a), and that several stellar evolutionary pathways may produce nebulae best catalogued as PNe. Current work suggests that PN-like nebulae may arise from the following channels:\\

\noindent $\bullet$ Post-AGB evolution of a single star (or member of a wide, non-interacting binary), producing a conventional or classical PN.\\
\noindent $\bullet$ Short-period binaries that have passed through a common-envelope (CE; De Marco 2009). Their PNe appear to have distinct observational properties (Bond \& Livio 1990; Frew \& Parker 2007; Miszalski et al. 2009b, 2010). Some post-CE PNe seem to contain two hot subdwarfs (Hillwig 2010), but the formation mechanism(s) of these, and the putative PNe around the classical novae V458~Vul and GK~Per remain unclear (Wesson et al. 2008; Rodr\'iguez-Gil et al. 2010).  Also noteworthy is the re-interpretation of V605~Aql (in Abell~58) by Lau, De Marco \& Liu (2010) as a possible nova in a PN.\\
\noindent $\bullet$ Longer-period interacting binaries, a subset of which appear to be related to the family of highly-collimated bipolar nebulae which host central stars exhibiting the B[e] phenomenon, with or without symbiotic characteristics.\\
\noindent $\bullet$ The so-called `born-again'  phenomenon, where a final helium flash produces a H-deficient star and surrounding knots inside a pre-existing, old PN (Zijlstra 2002), however there seem to be problems with this scenario (Lau et al. 2010).\\
\noindent $\bullet$ Scenarios that produce the rare O(He) stars; K 1-27 is the archetype of the class. They are possibly the long-sought successors of the R~CrB stars (Rauch et al. 2008), which may result in turn from a double-degenerate merging process (Clayton et al. 2007).\\
\noindent $\bullet$ Evolution of a super-AGB star (Poelarends et al. 2008), but there are no confirmed examples known.\\

There is also great variation in the spectral types of CSPN. Up to 20\% are H-deficient, including those where Wolf-Rayet features are present --- these are denoted [WR] to differentiate them from their Population I cousins. Almost all [WR] CSPN belong to the [WC]/[WO] sequence except for the rare [WN] objects N~66 in the LMC (Pe\~na et al. 1995; Hamann et al. 2003), PM~5 in the Galaxy (Morgan, Parker \& Cohen 2003), and two recently discovered Galactic CSPN belonging to a proposed new [WN/WC] class: PB~8 (Todt et al. 2010) and Abell~48 (DePew et al. 2010); the origin of these peculiar stars is unclear. N~66 surrounds a [WN4-5] star and has been suggested to be the product of a high-mass progenitor, but a binary evolution channel may be more likely (see Hamann et al. 2003, for a discussion).

\section{Relative Numbers}
It is difficult to ascertain the relative contributions of these various evolutionary pathways to the PN population until a complete spectroscopic survey of all nearby CSPN has been undertaken.  Based on a preliminary 1.0~kpc volume-limited sample, Frew \& Parker (2007) found 22 $\pm$ 9$\%$ of local CSPN are close binaries and are presumably derived from a post-CE evolutionary channel (De Marco 2009).  We find the fraction of [WR] CSPN to be 7 $\pm$ 3\%, and H-deficient CSPN to be about 20\% of the total.  Some nearby PNe possess unresolved high-density cores, which we have termed `EGB~6-like' CSPN (Frew \& Parker 2010a).  Two are known within 1.0~kpc, so the fraction is $\sim$2/55 or 4 $\pm$ 2$\%$. This is a lower limit, as not all local CSPN have adequate spectroscopy. 

Within a larger sample of 210 PNe within 2.0~kpc (Frew 2008), some rarer subclasses make an appearance.  There are but one or two born-again objects (Abell 30 and Abell 78), one PN with an O(He) nucleus (K~1-27), one putative PN with a [WN]~CSPN (PM~5), another with a [WN/WC] nucleus (Abell 48; DePew et al. 2010), and one or two highly collimated outflows associated with [Be] stars (which may not be true PNe), e.g. M~2-9.   More robust estimates of these relative proportions will follow from a more detailed analysis of our enlarged 2.0 and 3.0~kpc samples.

\section{Future Work}
The total number of Galactic PNe is now nearly 3000, with the potential to grow more over the next few years (Parker \& Frew 2010a; Jacoby et al. 2010).  New discoveries from IPHAS (Viironen et al. 2009; Sabin et al. 2010) plus new candidates found at mid-IR wavelengths (e.g. Mizuno et al. 2010) should substantially add to the total.  It is important that newly discovered nebulae be correctly classified, and it pays for the specialist to be cognisant of the full zoo of potential mimics.  Much improved discrimination of true PN from their mimics is now possible based on the range of online UV, optical, near-IR, mid-IR, and radio data-sets that are available.  While data-mining has its place, we caution that unusual objects be classified on a case-by-case basis.

One outstanding question from the two most recent APN meetings is the role of binarity in the formation and shaping of PNe (De Marco 2009, 2010), so we intend to carefully survey our samples for new close binaries to better ascertain the proportion of post-CE PNe in the local Galactic disk.  This number can then be confronted with model  predictions made by other authors (e.g. Soker 1997; Moe \& De Marco 2006).  Our preliminary disk post-CE estimate of $\sim$22$\%$ is in agreement with the estimate of 17 $\pm$ 5$\%$ for a large, but flux-limited sample of Galactic bulge CSPN (Miszalski 2009a).  We lastly note a surprising diversity of emission-line stars in PN-like nebulae, but the links between these nuclei and some symbiotic and B[e] stars is still unclear.  A complete spectroscopic survey of all local CSPN is needed to better quantify the observed diversity of spectra. In particular, multiwavelength follow-up of unusual CSPN is required to confront evolutionary theory, and in this sense, they are much more valuable than the more routine objects.  We look forward to the day when a precise taxonomy of the `PN phenomenon' is attained, but a consensus appears elusive at this point in time.

\acknowledgements 
DJF would like to thank the SOC for the opportunity to present an invited talk, and we thank Macquarie University for travel funding.

\section*{References}
{\small
Balick, B. \& Frank, A. 2002, ARA\&A, 40, 439\\
Bond, H.E. \& Livio, M. 1990, ApJ, 355, 568\\
Ciardullo, R. 2010, PASA, 27, 149\\
Clayton, G.C., Geballe, T.R., Herwig, F., Fryer, C. \& Asplund, M. 2007, ApJ, 662, 1220\\
Cohen, M.C., et al. 2007, ApJ, 669, 343\\
Cohen, M.C., et al. 2010, MNRAS, submitted\\
Corradi, R.L.M. 2003, ASPC, 303, 393\\
Corradi, R.L.M., et al. 2008, A\&A, 480, 409\\
De Marco, O. 2009, PASP, 121, 316\\
De Marco, O. 2010, APN5 Proceedings (Ebrary), eprint: arXiv:1010.3802\\
DePew, K., et al. 2010, MNRAS, submitted\\
Frew, D.J. 2008, PhD thesis, Macquarie University\\
Frew, D.J. \& Parker, Q.A. 2006, IAUS, 234, 49\\
Frew, D.J. \& Parker, Q.A. 2007, APN4 Proceedings, IAC Electronic Pub., p. 475\\
Frew, D.J. \& Parker, Q.A. 2010a, PASA, 27, 129\\
Frew, D.J. \& Parker, Q.A. 2010b, in preparation\\
Frew, D.J., Madsen, G.J., O'Toole, S.J., \& Parker Q.A. 2010a, PASA, 27, 203\\
Frew, D.J., et al. 2010b, PASA (in press), eprint: arXiv:1009.5914\\
Hamann, W.-R., Pe\~na, M., Gr\"afener, G. \& Ruiz, M.T. 2003, A\&A, 409, 969\\
Hillwig, T. 2010, APN5 Proceedings (Ebrary), eprint: arXiv:1010.1026 \\
Jacoby, G.H., et al. 2010, PASA, 27, 156\\
Lau, H.H.B., De Marco, O. \& Liu X.-W. 2010, MNRAS (in press), eprint: arXiv:1009.3138\\
Madsen, G.J., Frew, D.J., Parker, Q.A., Reynolds, R.J. \& Haffner L.M. 2006, IAUS, 234, 455\\
Miszalski, B., et al. 2008, MNRAS, 384, 525\\
Miszalski, B., Acker, A., Moffat, A.F.J., Parker, Q.A. \& Udalski, A. 2009a, A\&A, 496, 813\\
Miszalski, B., Acker, A., Parker, Q.A. \& Moffat, A.F.J. 2009b, A\&A, 505, 249\\
Miszalski, B., et al. 2010, APN5 Proceedings (Ebrary), eprint: arXiv:1009.2890\\
Mizuno, D.R., et al. 2010, AJ, 139, 1542\\
Moe, M. \& De Marco, O. 2006, ApJ, 650, 916\\
Morgan, D.H., Parker, Q.A. \& Cohen, M. 2003, MNRAS, 346, 719\\
Osterbrock, D.E. \& O'Dell, C.R. (eds) 1968, Planetary Nebulae, IAUS, 34 (Reidel: Dordrecht)\\
Parker, Q.A. \& Frew, D.J. 2010, APN5 Proceedings (Ebrary)\\
Parker, Q.A., et al. 2006, MNRAS, 373, 79\\
Pe\~na, M., Peimbert, M., Torres-Peimbert, S., Ruiz, M.T. \& Maza, J. 1995, ApJ, 441, 343\\
Poelarends, A.J.T., Herwig, F., Langer, N. \& Heger, A. 2008, ApJ, 675, 614\\
Rauch, T., Reiff, E., Werner, K. \& Kruk, J.W. 2008, ASPC, 391, 135\\
Rodr\'iguez-Gil, P., et al. 2010, MNRAS, 407, L21\\
Sabin, L., et al. 2010, in preparation\\
Sahai, R., Morris, M., S\'anchez Contreras, C. \& Claussen, M. 2007, AJ, 134, 2200\\
Schmeja, D. \& Kimeswenger, S. 2001, A\&A, 377, L18\\
Soker, N. 1997, ApJS, 112, 487\\
Szczerba, R., Si\'odmiak, N., Stasi\'nska, G. \& Borkowski, J. 2007, A\&A, 469, 799\\
Todt, H., Pe\~na, M., Hamann, W.-R. \& Gr\"afener, G. 2010, A\&A, 515, 83 \\
Viironen, K., et al. 2009, A\&A, 504, 291\\
Wesson, R., et al. 2008, ApJ, 688, L21\\
Zijlstra, A.A. 2002, Ap\&SS, 279, 171\\
}


\end{document}